\documentstyle[aps,preprint,version2]{revtex}

\tightenlines
\begin{document}

\preprint{ hep-th/9211076 }
\preprint{~}
\preprint{TUW-92-18}

\draft

\begin{title}
Thermal Green's Functions\\
{}From Quantum Mechanical Path Integrals
\end{title}

\author{D.G.C. McKeon\thanks{Internet address:
{\tt tmleafs@uwovax.uwo.ca}} }
\begin{instit}
Department of Applied Mathematics, University of Western Ontario,\\
London, Ontario, CANADA N6A 5B7
\end{instit}

\author{A. Rebhan\thanks{Internet address:
{\tt rebhana@email.tuwien.ac.at}} }
\begin{instit}
Institut f\"ur Theoretische Physik der Technischen Universit\"at Wien,\\
Wiedner Hauptstr. 8--10, A--1040 Vienna, AUSTRIA
\end{instit}

\begin{abstract}

In this paper it is shown how the generating functional for
Green's functions in relativistic quantum field theory and in
thermal field theory can be evaluated in terms of
a standard quantum mechanical path integral. With this
calculational approach one avoids the loop-momentum integrals
usually encountered in Feynman perturbation theory, although
with thermal Green's functions, a discrete sum
(over the winding numbers of paths
with respect to the circular imaginary time)
must be computed.
The high-temperature expansion of this sum can be performed
for all Green's functions at the same time, and is particularly
simple for the static case.
The procedure is illustrated by evaluating the two-point function
to one-loop order in a $\phi^3_6$ model.
\end{abstract}

\narrowtext
\section{Introduction}

Using Schwinger's proper-time representation \cite{r4},
the regulated generating functional for Green's functions in
relativistic quantum field theory and in thermal field theory
can be formulated in terms of the matrix elements
$\langle x|e^{-iHt}|y\rangle $ and
$\langle x|e^{-Ht}|y\rangle $, respectively, where $H$
is generally of the form $H={1\over2} (p-A)^2+V$. Usually these are
evaluated using the Hamiltonian formalism leading to Feynman
integrals for Green's functions.

On the other hand, in Ref.~\cite{r4} Schwinger has demonstrated how
in particular the one-loop effective action can be evaluated by
using quantum mechanical methods for the matrix elements appearing
in the proper-time formulation.
In Ref.~\cite{r6}, it has recently been shown by
one of the present authors
how these matrix elements can be computed by standard quantum mechanical
path integrals, and that these can be employed to perform perturbation
theory at arbitrary loop order, using the formalism of operator
regularization \cite{r3}. Independently, Strassler \cite{r9b}, motivated
by the techniques developed in string theory, has proposed an alternative
formulation based on quantum mechanical path integrals, restricted, however,
to one-loop order. First-quantized string theory is usually formulated
through path integrals over fields defined on two-dimensional world-sheets.
It therefore seems only natural to expect the possibility of rephrasing
ordinary field theory in terms of path integrals over functions
defined on one-dimensional world lines.
Indeed, such an approach has already been outlined by Polyakov
in Ref.~\cite{rP}. The parallel with string theory
can be drawn in particular for the one-loop effective action in
quantum field theory, where there are no internal vertices.
Nevertheless, as shown in Refs.~\cite{r6,r9a} in a number of explicit
examples, the calculational approach based on quantum mechanical
path integrals can be exploited also at higher loop order and for
arbitrary fields.

In the present paper, we shall show how this technique can be extended
to the evaluation of Green's functions in the imaginary-time formulation
of thermal field theories.
Again we find that loop-momentum integrals are obviated, however
because of the topological constraint from the circular imaginary
time, an infinite sum corresponding to the winding number of paths
in imaginary time appears. This sum is equivalent to but not
identical with the conventional sum over loop Matsubara frequencies.
In fact, we shall show that the representation as a
sum over winding numbers suggests
a novel method of evaluating the high-temperature expansion of
Green's functions. This expansion can be done for all Green's
functions at the same time, with the coefficients being integrals
over the parameters corresponding to the insertion of external vertices.
This turns out to be comparatively simple for the static (but
momentum-dependent) case.
The procedure is worked out for a $\phi^3$-model, and explicit results
are derived for the one-loop two-point function.

\newpage
\section{Loop diagrams without \\ loop-momentum integrals}

If we consider a field theoretical model with an action
\begin{equation}
S[\phi]=\int d^Dx \left[
-{1\over2}(\partial _\mu\phi)^2-{1\over2}m^2\phi^2
-{1\over 3!}\lambda \phi^3 \right],
\label{S}\end{equation}
(metric signature $(-,+,\ldots,+)$ )
then by introducing a background field \cite{r2} $f(x)$
the generating functional for one-particle irreducible Green's
functions can be formulated in terms of the operator
\begin{equation}
H={1\over2}\left[p^2+m^2+\lambda f\right],\label{H}
\end{equation}
where $p=-i\partial$.

The one-loop generating functional is given by
\begin{equation}
i\Gamma^{(1)}[f]={1\over2}\int dx \langle x|\ln H|y\rangle,
\end{equation}
whereas at higher loop orders $\langle x|H^{-1}|y \rangle$ is
the basic building block.

In operator regularization \cite{r3} both $\ln H$ and $H^{-1}$
are defined through derivatives of $H^{-s}$,
\begin{mathletters}
\begin{eqnarray}
\ln H&=&\lim_{s\to0}-{d\over ds}H^{-s},\\
H^{-1}&=&\lim_{s\to1}{d\over ds}(s-1)H^{-s},
\end{eqnarray}
\end{mathletters}
which is represented as
\begin{equation}
H^{-s}={1\over\Gamma(s)}\int_0^\infty dit(it)^{s-1}e^{-iHt}.
\end{equation}

In this way the regulated generating functional (which turns out
to be finite upon taking the limit in $s$) is expressed entirely
in terms of matrix elements $\langle x|\exp -iHt|y\rangle$;
the one-loop generating functional in particular is derived from
the $\zeta$-function \cite{r1}
\begin{eqnarray}
\zeta_H(s)&=&{1\over \Gamma(s)} \int_0^\infty dit(it)^{s-1}\nonumber\\
&&\times
\int dx\,dy\,
\delta^D(x-y) \langle x|\exp -iHt|y\rangle ,\label{z}
\end{eqnarray}
according to
\begin{equation}
i\Gamma^{(1)}[f]={1\over2}\lim_{s\to0}{d\over ds}\zeta_H(s).
\label{G1}\end{equation}
(Operator regularization at higher loop orders is discussed in
Ref.~\cite{r5}.)

The perturbative evaluation can proceed by using the Schwinger
expansion \cite{r4}
\FL
\begin{eqnarray}
e^{-i(H_0+H_1)t}&=&e^{-iH_0t}-it\int_0^1 du\,e^{-i(1-u)H_0t}H_1
e^{-iuH_0t}\nonumber\\&&+\ldots
\end{eqnarray}
and inserting complete sets of states, which involves having
to compute loop-momentum integrals.

An alternative \cite{r6}
for the evaluation of the matrix elements
$\langle x|\exp-iHt|y\rangle$
is to use the standard quantum mechanical
path integral \cite{r7}
\begin{eqnarray}
&&\langle x|\exp -i\left[{1\over2}(p-A)^2+V\right]t|y\rangle
\nonumber\\
&&= \int Dq(\tau)\exp i\int_0^t d\tau
\biggl[{1\over2}\dot q^2(\tau)\nonumber\\
&&\qquad\qquad+\dot q(\tau)\cdot A(q(\tau))-V(q(\tau))\biggr]
\label{pi}\end{eqnarray}
with $q(0)=y$, $q(t)=x$.
Using a perturbative expansion of the right hand side of
Eq.~(\ref{pi}),
we can write for the matrix element $\langle x|\exp -iHt|y\rangle$
with $H$ given by Eq.~(\ref{H})
\begin{eqnarray}
M_{xy} &\equiv&
\langle x|\exp -{i\over 2}\left[p^2+\lambda f\right]t|y\rangle
\nonumber\\
&=& \int Dq(\tau) \sum_{N=0}^\infty \left({-i\lambda\over 2}\right)^N
\prod_{i=1}^N \int_0^t d\tau_i f(q(\tau_i))\nonumber\\
&&\times\exp i\int_0^t d\tau{\dot q^2(\tau)\over 2},
\label{M}\end{eqnarray}
where we have omitted the trivial factor $\exp-i(m^2/2)t$.
Upon Fourier decomposition of the background field,
\FL
\begin{equation}
f(q(\tau_i))={1\over (2\pi)^{D/2}}\int d^Dk_i \tilde f(k_i)
\exp-ik_i\cdot q(\tau_i),
\label{f}\end{equation}
this matrix element can be expressed in terms of the
standard result \cite{r7,r8} for a particle subject to an
external force
\begin{mathletters}
\begin{eqnarray}
&&\int Dq(\tau)\exp i\int_0^t d\tau
\left[{1\over2}\dot q^2(\tau)-\gamma_N(\tau)\cdot q(\tau)\right]
\nonumber\\
&=&{1\over (2\pi it)^{D/2}}\exp i\biggl[
{(x-y)^2\over 2t}\nonumber\\
&&-{1\over t}\int_0^t d\tau \gamma_N(\tau)\cdot
\left[x\tau+y(t-\tau)\right] \nonumber\\
&&+{1\over2}\int_0^t d\tau'd\tau''
\gamma_N(\tau')\cdot\gamma_N(\tau'')G(\tau',\tau'') \biggr]
\label{pig}\end{eqnarray}
with
\begin{equation}
\gamma_N(\tau)=\sum_{i=1}^N \delta(\tau-\tau_i) k_i
\end{equation}
and $G$ being the one-dimensional Green's function
\begin{equation}
G(\tau,\tau')=
{1\over2}|\tau-\tau'|-{1\over2}(\tau+\tau')+{\tau\tau'\over t}.
\label{G}\end{equation}
\end{mathletters}

By Eq.~(\ref{pig}), the contribution to
Eq.~(\ref{M}) that is proportional
to $\lambda^N$ when $x=y$, which determines the one-loop
$N$-point Green's function, is
\begin{eqnarray}
&&M_{xx}^{(N)}={1\over N!}\left({-i\lambda\over 2(2\pi)^{D/2}}\right)^N
\int\prod_{i=1}^N d^Dk_i \tilde f(k_i) \nonumber\\
&&\quad\times
\int_0^t d\tau_1\ldots d\tau_N {1\over (2\pi it)^{D/2}}\nonumber\\
&&\quad\times\exp i\left[ \sum_{i=1}^N k_i\cdot x +{1\over2}\sum_{i,j=1}^N
k_i\cdot k_j G(\tau_i,\tau_j) \right].
\label{MxxN}\end{eqnarray}
When substituted into Eq.~(\ref{z}), the integral over $x$
implements momentum conservation ($\sum_i k_i=0$), and the
integration variables $\tau_i$,
appropriately combined and rescaled, turn out to correspond to
the usual Feynman parameters. At no stage, however, does one
encounter any loop-momentum integral.

This procedure can be applied beyond one-loop order \cite{r9a}
and to models involving spinors and vectors \cite{r9a,r9b}.
An advantage of using this formalism when dealing with a
non-Abelian vector field theory is that no complicated three or
four point vertices need be handled, as can be seen from Ref.~\cite{r3}.

\section{Thermal Green's functions}

We now show how this technique can be used to evaluate thermal
Green's functions. If one were to determine these quantities
associated with the model of Eq.~(\ref{S}), a starting point is
the partition function \cite{r10}
\FL
\begin{eqnarray}
Z&=&\int_{periodic}\!\!\!\!\!\!\!\!\!\!\!\!\!
D\phi\;\exp\biggl\{\int_0^\beta dx_0\int d^{D-1}x \nonumber\\
&\times&\Bigl(-{1\over2} \left[
\left({\partial \phi\over \partial x_0}\right)^2
+\left(\nabla\phi\right)^2+m^2\phi^2\right]
-{\lambda\over 3!}\phi^3 \Bigr)\biggr\},
\label{Z}\end{eqnarray}
where ``periodic'' denotes the constraint
that $\phi$ is periodic in imaginary time $x_0$,
\begin{equation}
\phi(x_0,{\bf x})=\phi(x_0+\beta,{\bf x}).
\label{periodic}\end{equation}

The regulated generating functional for one-loop Green's functions
is again determined by a $\zeta$-function analogous to
Eq.~(\ref{z}),
\begin{eqnarray}
\zeta_H(s)&=&{1\over \Gamma(s)} \int_0^\infty dt\,t^{s-1}\nonumber\\
&&\times\int dx\,dy\,
\delta^D(x-y) \langle x|\exp -Ht|y\rangle ,\label{zb}
\end{eqnarray}
where the $p^2$ contained in $H$ is
contracted with a Euclidean metric.

The central matrix element to be evaluated and its path integral
representation now are
\FL
\begin{eqnarray}
M^\beta_{xy}&=&
\langle x|\exp-{1\over2}\left[p^2+\lambda f\right]t|y\rangle \nonumber\\
&=& \int Dq(\tau)\exp-\int_0^t d\tau
{1\over2}\left[\dot q^2(\tau)+\lambda f(q(\tau))\right]
\label{Mb}\end{eqnarray}
with $q(0)=y$, $q(t)=x$, and with the identification
\begin{equation}
q_0\equiv q_0+n\beta, \qquad n \;\hbox{integer}.
\label{q0b}\end{equation}

Paralleling the steps of the previous section, one can
again perform a perturbative expansion in $\lambda$ and restrict
oneself to plane waves
\FL
\begin{equation}
\sqrt\beta (2\pi)^{D-1\over 2}f(q(\tau_i))=
\exp-i\left(\omega_{n_i} q_0(\tau_i)+{\bf k}_i\cdot{\bf q}(\tau_i)\right),
\label{fb}\end{equation}
where the periodicity condition
(\ref{periodic}) restricts $\omega_{n_i}$ to the discrete
set of Matsubara frequencies
\begin{equation}
\omega_n={2\pi n\over \beta}.
\label{omegan}\end{equation}
An $N$-point Green's function is thus determined by the
expression
\begin{mathletters}
\FL
\begin{eqnarray}
&&M^{\beta(N)}_{xy} \nonumber\\
&&=
{1\over N!}\left(-\lambda\over 2(2\pi)^{(D-1)/2}\sqrt\beta\right)^N
\!\!\int_0^t d\tau_1\ldots d\tau_N P^\beta_{xy}[\gamma_N],\label{MN}\\
&&P^\beta_{xy}[\gamma_N] \nonumber\\
&&=\int Dq(\tau)\exp-\int_0^t d\tau
\left[{1\over2}\dot q^2(\tau)-\gamma_N(\tau)\cdot q(\tau)\right],
\label{Pbg}\end{eqnarray}
where
\begin{eqnarray}
i\gamma_N^0(\tau)&=&\sum_{i=1}^N \omega_{n_i}^0 \delta(\tau-\tau_i),
\nonumber\\
i\vec\gamma_N(\tau)&=&\sum_{i=1}^N {\bf k}_i \delta(\tau-\tau_i).
\label{igamma}\end{eqnarray}
\end{mathletters}

However, because of the topological constraint (\ref{q0b}),
it is no longer possible to perform a shift of variable
\begin{equation}
q(\tau)=\tilde q(\tau)-\int_0^t d\tau' G(\tau,\tau')
\gamma_N(\tau')
\label{qshift}\end{equation}
with $G$ given by Eq.~(\ref{G})
to bring the path integral (\ref{Pbg})
in Gaussian form. Instead of a free particle
subject to an external force,
the quantum mechanical problem
described by the path integral (\ref{Pbg})
is now that of a particle on a circle
(with respect to $q_0$). Its path integral treatment
has been given in Ref.~\cite{r11},
and the solution, adapted to our case, reads
\begin{equation}
P^\beta_{xy}=\sum_{n=-\infty}^\infty P^\infty_{(x_0+n\beta,{\bf x})y}.
\label{Pbsum}\end{equation}
This corresponds to summing over the paths from $y$ to $x$
differing in winding number $n$ around the circular imaginary time.

We consequently need only compute a zero-temperature ($\beta=\infty$)
path integral with $q(0)=y$, ${\bf q}(t)={\bf x}$, and $q_0(t)=x_0+n\beta$,
then perform the sum over $n$, in order to determine the
finite-temperature path integral (\ref{Pbg}). With $\beta=\infty$,
the shift of variable (\ref{qshift}) is legitimate.

\widetext
Upon substituting Eq.~(\ref{qshift}) into
Eq.~(\ref{Pbg}) at $\beta=\infty$, we find that
\FL
\begin{eqnarray}
P^\infty_{xy}[\gamma_N]&=&
\exp\biggl[{1\over t}\int_0^t d\tau
\left[x\tau+y(t-\tau)\right]\cdot\gamma_N(\tau) 
-{1\over2}\int_0^t d\tau'd\tau''\gamma_N(\tau')\cdot\gamma_N(\tau'')
G(\tau',\tau'') \biggr] \nonumber\\
&&\times
\int D\tilde q(\tau)\exp-\int_0^t d\tau{1\over2}\dot{\tilde q}^2(\tau).
\label{pis}\end{eqnarray}
The remaining path integral is given by
\begin{eqnarray}
\int Dq(\tau)\exp-\int_0^t d\tau{1\over2}\dot q^2(\tau)
&=& \langle x|\exp-{p^2\over 2}t|y\rangle  \nonumber\\
=\int {d^Dk\over (2\pi)^D} e^{ik\cdot(x-y)-k^2t/2}
&=& {1\over (2\pi t)^{D/2}}e^{-(x-y)^2\over 2t}.
\label{pigauss}\end{eqnarray}

By Eqs.~(\ref{Pbsum}), (\ref{pis}), and (\ref{pigauss}), we obtain
\begin{eqnarray}
M^{\beta(N)}_{xy}
&=&
{1\over N!}\left(-\lambda t\over 2(2\pi)^{(D-1)/2}\sqrt\beta\right)^N
\int_0^1 d\sigma_1\ldots d\sigma_N \sum_{n=-\infty}^\infty
{1\over (2\pi t)^{D/2}} \nonumber\\
&&\times\exp\biggl\{-\left[(x-y)^2+2n\beta(x_0-y_0)
+n^2\beta^2\right]/2t \nonumber\\
&&\qquad
-i(x-y)\cdot\sum_{i=1}^N k_i\sigma_i
-i y\cdot\sum_{i=1}^N k_i-in\beta\sum_{i=1}^N \omega_{n_i}\sigma_i
\nonumber\\
&&\qquad +{1\over2}\sum_{i,j=1}^N
\left({\bf k}_i\cdot{\bf k}_j+\omega_{n_i}\omega_{n_j}\right)
G(\sigma_i t,\sigma_j t) \biggr\},
\label{MNf}\end{eqnarray}
where we have rescaled $\tau_i=\sigma_i t$.

The sum over $n$ can be expressed in terms of the Jacobi function
\begin{equation}
\theta_3(\nu|\tau)=\sum_{n=-\infty}^\infty\exp i\pi\left[\tau n^2
+2\nu n\right],
\label{th3}\end{equation}
so that
\begin{eqnarray}
M^{\beta(N)}_{xy}&=&
{1\over N!}\left(-\lambda t\over 2(2\pi)^{(D-1)/2}\sqrt\beta\right)^N
\int_0^1 d\sigma_1\ldots d\sigma_N {1\over (2\pi t)^{D/2}} \nonumber\\
&&\times\exp\biggl\{-{(x-y)^2\over 2t}-i\sum_{i=1}^N
\left[\sigma_i(x-y)+y\right]\cdot
 k_i 
+{1\over2}\sum_{i,j=1}^N
 k_i\cdot k_j G(\sigma_i t,\sigma_j t) \biggr\}
\nonumber\\&&\times
\theta_3\left({i\beta\over 2\pi}
[{x_0-y_0\over t}+i\sum_{i=1}^N \sigma_i
\omega_{n_i}] \bigg| {i\beta^2\over  2\pi t}\right).
\label{MNth}\end{eqnarray}

\narrowtext
At one-loop order, only the matrix element with $x=y$ is needed,
and integration over $x$ again enforces $\sum_i k_i=0$.
In particular, in $D=6$ dimensions, the two-point function
with ${\bf k}_1={\bf k}=-{\bf k}_2$ and
$\omega_{n_1}=\omega_n=-\omega_{n_2}$
is determined by
\FL
\begin{eqnarray}
&&\zeta^{(2)}(s)={1\over \Gamma(s)}
{\lambda^2\over 64\pi^2 \beta}\int_0^\infty dt\,
t^{s-2} \int_0^1 d\sigma_1 d\sigma_2 \nonumber\\
\times&&
\exp\left\{
\left[(\omega_n^2+{\bf k}^2)
(-|\sigma_1-\sigma_2|+(\sigma_1-\sigma_2)^2)
-m^2\right]{t\over 2}
\right\} \nonumber\\
\times&&
\theta_3\left({-\beta\over 2\pi}(\sigma_1-\sigma_2)\omega_n
\bigg|{i\beta^2\over 2\pi t}\right).
\label{z2}\end{eqnarray}
The symmetric form of the integrand together with the
periodicity of the $\theta_3$-function
allows us to simplify $\int_0^1 d\sigma_1 d\sigma_2
\to \int_0^1 d(\sigma_1-\sigma_2)$ in Eq.~(\ref{z2}).
At zero temperature (where $\theta_3\to1$), $u=\sigma_1-\sigma_2$
would exactly correspond to the usual Feynman parameter for
combining two propagators. The sum defining $\theta_3$, however,
does not correspond to the conventional sum over loop Matsubara
frequencies. The latter are recovered after employing
Jacobi's imaginary transformation \cite{Erd}
\begin{mathletters}
\begin{equation}
\theta_3\left({\nu\over \tau}\bigg|-{1\over \tau}\right)=
(-i\tau)^{1\over2} e^{i\pi\nu^2/\tau}\theta_3(\nu|\tau),
\label{Jit}\end{equation}
which in our case reads
\begin{eqnarray}
&&\theta_3\left(-{\beta\over 2\pi}\sum_i \sigma_i\omega_{n_i}\bigg|
{i\beta^2\over 2\pi t}\right)\nonumber\\
&=&{\sqrt{2\pi t}\over \beta}
\theta_3\left({i\over \beta}\sum_i \sigma_i\omega_{n_i} t\bigg|
{2\pi i t\over \beta^2}\right)\nonumber\\&&
\times\exp\Bigl\{-{1\over2}(\sum_i \sigma_i\omega_{n_i})^2 t\Bigr\}.
\label{Jit2}\end{eqnarray}
\end{mathletters}
With Eq.~(\ref{Jit2}), Eq.~(\ref{z2}) coincides with the
conventional expression for the finite-temperature
two-point Green's function where the loop-momentum integration
has been done, identifying $2\pi n/\beta$ with
the usual loop Matsubara frequency. It should be noticed that no loop
momentum integral was ever encountered in the derivation of
Eq.~(\ref{z2}); the sum over loop frequencies is
replaced by the equivalent but conceptually different sum
over winding numbers in Eq.~(\ref{Pbsum}).\footnote{The difference
of these two sums would have been more conspicuous, had we
considered fermionic fields:
there we would have had an alternating sum over integer
winding numbers on the one hand, and on the other a sum
over Matsubara frequencies determined by half-integers.
}

\section{High-temperature expansion}

The standard way to evaluate the finite-temperature Green's
functions in the imaginary-time formalism \cite{r10} is to
do the sum over loop Matsubara frequencies first, keeping
the loop-momentum integrals as such, without combining denominators
by Feynman parametrization. On the other hand, in our calculational
approach, the loop-momentum integrals are circumvented,
leaving us with integrals equivalent to Feynman parameter integrals and
either with the sum over loop Matsubara frequencies or
the sum over winding numbers, which are Jacobi transforms of
each other.

It has recently been shown in Ref.~\cite{BFT}
how to obtain a systematic high-temperature expansion of
the sum over Matsubara frequencies after the other integrals
are all performed. In the following we shall demonstrate
that the representation as a sum over winding numbers,
which naturally emerges from the computational scheme
considered here,
is an interesting alternative as concerns the task of
a high-temperature expansion.
Keeping the integrations over the vertex insertion parameters
$\sigma_i$ to the end, one can in fact aim at deriving
a generic form of the high-temperature expansions of
all (one-loop) $N$-point Green's functions at the same time.
This will turn out to be feasible and comparatively simple, too, for the
static case $\omega_{n_i}=0$ (${\bf k}_i \not\equiv 0$).

Inserting the matrix element
(\ref{MNth}) into the $\zeta$-function representation of
a one-loop $N$-point Green's function, in the static case we are led to
compute the proper-time integral
\FL
\begin{eqnarray}
&&I_N = \int_0^\infty dt\,t^{s-1+N-D/2}
\theta_3\left(0 \Big| {i\beta^2\over  2\pi t}\right)
\exp\left[-{1\over2}Q_N t\right] \nonumber\\
&&= \sum_{n=-\infty}^\infty
\int_0^\infty dt\,t^{s-1+N-D/2}
\exp\left[-{n^2\beta^2\over2t}-{1\over2}Q_N t\right],
\label{IN}\end{eqnarray}
where
\FL
\begin{equation}
Q_N\equiv m^2-{1\over2}\sum_{i,j=1}^N {\bf k}_i\cdot{\bf k}_j
\left(|\sigma_i-\sigma_j|-
(\sigma_i+\sigma_j)+2\sigma_i \sigma_j\right)
\label{Q}\end{equation}
contains the dependence on external momenta and vertex insertion
parameters. Integration of the latter is postponed for the moment.

The integral in Eq.~(\ref{IN}) is readily performed \cite{GR},
and the sum naturally splits into a zero-temperature contribution
(the term with $n=0$) and temperature contributions,
\begin{eqnarray}
I_N &\equiv& I_N^\infty+I_N^\beta \nonumber\\
&=& (Q_N/2)^{\nu-s}\Gamma(s-\nu)\nonumber\\&&+
4\sum_{n=1}^\infty \left( {\sqrt{Q_N}\over n\beta} \right)^\nu
K_\nu(n\beta\sqrt{Q_N}),
\label{INb}\end{eqnarray}
with $\nu\equiv D/2-N$ and $K_\nu$ being a modified Bessel function
of integer (half-integer) order for even (odd) dimension $D$.
Since the temperature contribution is free of ultraviolet divergences,
we have put the regulating parameter $s=0$ there; the otherwise
necessary differention with respect to $s$ in Eq.~(\ref{G1})
just amounts to dropping the overall factor $1/\Gamma(s)$ in $\zeta_H$,
Eq.~(\ref{zb}).

The low-temperature expansion of $I_N^\beta$ in the form (\ref{INb})
can be obtained from
the asymptotic formula $K_\nu(z)\sim \sqrt{\pi/2z}\,e^{-z}$;
the sum over $n$ is, however,
slowly converging for high temperatures.
A series expansion in powers (and logarithm) of $\beta$ can be
found by the Mellin summation technique \cite{Mellin}, leading to
\FL
\begin{eqnarray}
&&I_N^\beta={1\over2\pi i}\int_{c-i\infty}^{c+i\infty}dz\,\zeta(z)
\nonumber\\&&\qquad\times
4\int_0^\infty dn\, \beta^{-\nu}Q_N^{\nu/2}n^{z-\nu-1}
K_\nu(\beta n\sqrt{Q_N}) \nonumber\\
&=&{1\over2\pi i}\int_{c-i\infty}^{c+i\infty}dz\,\zeta(z)\,
2^{z-\nu} \beta^{-z} Q_N^{\nu-z/2}\Gamma({z\over2})
\Gamma({z\over2}-\nu),
\label{intc}\end{eqnarray}
where $c$ is a real number with $c>{\rm max}(1,2\nu)$ and $\zeta(z)$
is the Riemann $\zeta$-function. Whereas the series representation
(\ref{INb}) is defined only for $\nu>-{1\over2}$, the integral
representation (\ref{intc}) allows us to analyticly continue to
general $\nu$.

In the form (\ref{intc}), $I_N^\beta$ can be evaluated
by closing the contour to the left and computing a sum over
residues. There is no contribution from the contour integral
over the large arc if $\beta\sqrt{Q_N}\ll 1$, which is
fulfilled in the limit of high temperature, $\beta\to0$.
For even dimension (integer $\nu$) we find
\widetext
\FL
\begin{eqnarray}
I_N^\beta &=&
\sum_{j=0}^{\nu-1} (-1)^j 2^{\nu-2j+1} {\Gamma(\nu-j)\over\Gamma(j+1)}
\zeta(2\nu-2j) \beta^{-2\nu+2j} Q_N^j 
+2^{1-\nu}\Gamma({1\over2})\Gamma({1\over2}-\nu)\beta^{-1}Q_N^{\nu-1/2}
\nonumber\\
&+&{2(-Q_N/2)^\nu \over \Gamma(\nu+1)}
\left[ \ln{\beta \sqrt{Q_N}\over4\pi}
-{1\over2}(\psi(1)+\psi(1+\nu))\right]
\nonumber\\
&+&2(-Q_N/2)^\nu \sum_{j=1}^\infty (-1)^j
{\Gamma(2j+1)\zeta(2j+1)\over\Gamma(j+1)\Gamma(j+1+\nu)}
\left({\beta\sqrt{Q_N}\over4\pi}\right)^{2j},\label{Iresult}
\end{eqnarray}
\narrowtext
where $\psi$ is the Digamma function. For nonpositive $\nu$ the
first sum does not contribute; for $\nu<0$, the logarithmic term
disappears, too, and the last sum starts at $j=|\nu|$.

A particular $N$-point Green's function is
determined by
$$\int_0^1 d\sigma_1\ldots d\sigma_N I_N.$$

For example, the two-point function in six dimensions ($\nu=1$)
is given by (dropping the zero-temperature part)
\begin{equation}
\Pi^\beta(\omega_n=0,{\bf k})={\lambda^2\over64\pi^3}\int_0^1 d\sigma_1
d\sigma_2 I_2^\beta,
\end{equation}
which in the massless case involves only elementary integrations,
yielding
\FL
\begin{eqnarray}
&&\Pi^\beta(\omega_n=0,{\bf k})\nonumber\\
&=&{\lambda^2\over128\pi^3}\biggl[
{4\pi^2\over3}\beta^{-2}-{1\over2}\pi^2 k \beta^{-1}
-{1\over3}k^2 \left( \ln{\beta k\over4\pi}+\gamma-{4\over3}\right)
\nonumber\\
&&-k^2\sum_{j=1}^\infty (-1)^j {\zeta(2j+1)\over(2j+1)(2j+3)}
\left({\beta k\over4\pi}\right)^{2j}\; \biggr].
\end{eqnarray}

Our results are a direct generalization of the high-temperature
expansions obtained in Refs.~\cite{Br,HW} for space-time
independent quantities such as thermodynamic potentials.
In fact, the same kind of infinite sums occurs there,
the essential difference being that here we still have to
evaluate the integrals over vertex insertion (or Feynman) parameters.
{Indeed,
the effective potential is contained in our results
as the special case ${\bf k}_i=0$, where $Q_N\equiv m^2$, independent of
the parameters $\sigma_i$, rendering the remaining integrals trivial.}

In the nonstatic case, however, where one could derive a generalized
version of Eq.~(\ref{intc}),
the comparative simplicity of the above derivation is lost.
There the condition $\beta\sqrt{Q_N}\ll 1$, which allowed us to
close the integration contour in Eq.~(\ref{intc}) to obtain
Eq.~(\ref{Iresult}) is not fulfilled
because then $\sqrt{Q}\sim \omega_n
\sim 1/\beta$.
It would be wrong to invoke
analytic continuation at this point and to substitute
$i\omega_n\to k_0+i\epsilon$,
with $k_0$ a continuous variable independent of $\beta$,
thereby dropping the contributions from the large arc after all.
This supposed transition to the real-time formulation
is problematic because Eq.~(\ref{intc}) has still to be
integrated over the Feynman parameters implicit in $Q_N$.
It has recently been shown by Weldon \cite{Wpre} that although
in the imaginary-time formulation 
Feynman-parametrized Green's functions are well-defined, they
have to be modified in the real-time theory.\footnote{The
uncorrected formulae of Feynman parametrization in the
real-time theory as
in Refs.~\cite{GHBD} correspond to an incorrect analytic
continuation which has a deceptive analytic behavior at
$k_0=0={\bf k}$, but possesses branch points at unphysical
locations, and is singular at infinity \cite{Wpre}.
The same phenomenon would occur by dropping the contribution
from the large arc after closing the contour in Eq.~(\ref{intc}).}

Let us emphasize that this presents no problem in principle
as our formalism is the imaginary-time one, and
we have verified equivalence to the usual results above. One has to be
careful, however, in performing the analytic continuation to
real frequencies. The latter is straightforward only with
the final result at hand, i.e.~after all the integrals have
been evaluated.

To summarize, we have shown that the recently proposed method
of evaluating loop diagrams in field theory by the use of
quantum mechanical path integrals can be extended to the
evaluation of thermal Green's functions in the imaginary-time
formalism. This alternative approach leads to a representation
of the latter in terms of sums over winding numbers
of paths 
in place of the
usual loop Matsubara frequencies. This different but
equivalent representation was shown to also suggest
alternative methods in performing high-temperature expansion,
and we have been able to derive a universal expansion for
static one-loop Green's functions.

The extension of our analysis to more interesting field theories
is mostly straightforward. In Refs.~\cite{r6,r9a} it has been
shown how for example Yang-Mills theory can be reformulated
in terms of quantum mechanical path integrals, if a Feynman-type
background field gauge is used, leading to a very economical
calculational scheme for the evaluation of Green's functions.
It is not equally
straightforward, though, to employ gauges that lead to a more complicated
kinetic term. Further, with operator regularization it is
almost mandatory to use a background-covariant gauge fixing \cite{AR}.
A somewhat more involved case is Green's functions
with external fermionic lines, because the different boundary
conditions of fermions and bosons have then to be accommodated
in one path integral expression. We intend to return to this
subject in a forthcoming publication.

\acknowledgements

One of us (D.G.C.M.) would like to thank NSERC for financial
support. R.~and D.~MacKenzie provided stimulus for this
investigation.


\end{document}